\documentclass[]{rmf-d}
\usepackage{rmfbib,multicol,times,epsf,amsmath,amssymb} 
\usepackage[latin1]{inputenc} 
\def\rmfcornisa{
}
\def\rmfcintilla{
}
\begin{document}
\input{feynman}
\newcommand{\ssc}{\scriptscriptstyle}
\newcommand{\be}{\begin{equation}}
\newcommand{\ee}{\end{equation}\noindent}
\newcommand{\bear}{\begin{eqnarray}}
\newcommand{\ear}{\end{eqnarray}\noindent}
\newcommand{\no}{\noindent}
\date{}
\renewcommand{\theequation}{\arabic{equation}}
\renewcommand{\arraystretch}{2.5}
\newcommand{\GeV}{\mbox{GeV}}
\newcommand{\cL}{\cal L}
\newcommand{\D}{\cal D}
\newcommand{\Dhalf}{{D\over 2}}
\newcommand{\Det}{{\rm Det}}
\newcommand{\PP}{\cal P}
\newcommand{\G}{{\cal G}}
\newcommand{\e}{{\rm e}}
\newcommand{\non}{\nonumber}
\def\R{1\!\!{\rm R}}
\def\Eins{\mathord{1\hskip -1.5pt
\vrule width .5pt height 7.75pt depth -.2pt \hskip -1.2pt
\vrule width 2.5pt height .3pt depth -.05pt \hskip 1.5pt}}
\newcommand{\symb}{\mbox{symb}}
\renewcommand{\arraystretch}{2.5}
\newcommand{\slD}{\raise.15ex\hbox{$/$}\kern-.57em\hbox{$D$}}
\newcommand{\slpartial}{\raise.15ex\hbox{$/$}\kern-.57em\hbox{$\partial$}}

\title{Multiloop Information from the QED Effective Lagrangian}
\vspace{-6pt}
\author{Gerald V. Dunne}
\address{Department of Physics and Department of Mathematics\\
University of Connecticut\\
Storrs CT 06269, USA}
\author{\underline {Christian Schubert}}
\address{Department of Physics and Geology
\\
University of Texas Pan American
\\
Edinburg, TX 78541-2999, USA
\\
schubert@panam.edu}

\maketitle


\vspace{20pt}
\begin{center}
Talk given by C.S. at {\it IX Mexican Workshop on Particles and Fields},
Colima, Mexico, November 17-22, 2003\\
(to appear in the conference proceedings)
\end{center}

\vspace{20pt}

\begin{abstract}
We obtain information on the QED photon amplitudes at high orders
in perturbation theory starting from known results on the
QED effective Lagrangian in a constant electric field. 
A closed-form all-order result for the weak field limit of the imaginary part of this 
Lagrangian has been given years ago by  Affleck, Alvarez and Manton (for scalar QED) and 
by Lebedev and Ritus (for spinor QED). We discuss the evidence for its 
correctness, and conjecture an analogous formula for the
case of a self-dual field. From this extension we then obtain, using Borel
analysis, the leading asymptotic growth for large $N$ of the maximally
helicity violating component of the $l$ - loop $N$ - photon amplitude in the low energy limit. 
The result leads us  to conjecture that the perturbation series
converges for the on-shell renormalized QED $N$ - photon amplitudes
in the quenched approximation. 
\end{abstract}
\keys{Quantum electrodynamics; perturbation series; Euler-Heisenberg Lagrangian; photon amplitudes. \vspace{-4pt}} \pacs{11.15.Bt;11.15.Tk;12.20.Ds \vspace{-4pt}}
\begin{multicols}{2}

\section{Introduction: The perturbation series in quantum electrodynamics}
\label{introduction}
\renewcommand{\theequation}{1.\arabic{equation}}
\setcounter{equation}{0}
In quantum electrodynamics, physical quantities are usually computed as a perturbative
series in powers of the fine structure constant $\alpha = {e^2\over 4\pi}$:

\bear
F(e^2) = c_0 + c_2 e^2 + c_4 e^4 + \ldots 
\label{pt}
\ear
In the early years of quantum field theory it was hoped that this type of
series would turn out to be convergent in general, or at least 
for sufficiently small values of the coupling constant(s). However, in 1952 Dyson \cite{dyson}
argued on physical grounds that QED perturbation theory should be divergent. 
Dyson argues: ``Suppose, if possible, that the series (1.1) converges for some positive value of $e^2$; this implies that $F(e^2)$ is an analytic function of $e$ at $e=0$. Then for sufficiently small values of $e$, $F(-e^2)$ will also be a well-behaved analytic function with a convergent power-series expansion'' \cite{dyson}. He then argues physically that this cannot be the case,
since for $e^2<0$ the QED vacuum
will be unstable due to a runaway production of $e^+e^-$ pairs which coalesce into
like-charge groups. 
This argument does not prove divergence or convergence, but it gives a first 
hint of the connection between
perturbative divergence and instability in quantum field theory
\cite{LeGuillou,Stevenson,Fischer,minneapolis}.

To substantiate this divergence mathematically in the case of QED  
turns out to be exceedingly difficult \cite{LeGuillou}.
Explicit all-order calculations are, of course, next to impossible, so that the natural
thing to try is to prove the divergence by establishing lower bounds on the contributions
of the individual Feynman diagrams representing the coefficients of the 
perturbation series. However this is still very hard to do for QED diagrams since here
the integrands have no definite sign. On the other hand, in scalar field theories the integrands 
have a definite sign (in the Euclidean)
so that here lower bounds can be established using inequalities such as

\bear
\prod_{i=1}^F \biggl({1\over p_i^2 + \kappa^2}\biggr)
\geq
{F^F\over \Bigl(\sum_{i=1}^Fp_i^2 + F\kappa^2\Bigr)^F}
\label{inequ}
\ear
At about the same time that Dyson's paper appeared, Hurst \cite{hurst} and Thirring \cite{thirring}
(see also Petermann \cite{petermann}) used this lower bound strategy to show by explicit
calculation that perturbation theory diverges in scalar $\lambda\phi^3$ theory for {\sl any}
value of the coupling constant $\lambda$. Hurst's proof is quite
straightforward; its essential steps are (i) the use of (\ref{inequ}) in the parametric representation
to find lower bounds for arbitrary diagrams (ii) a proof that the number of distinct Feynman
diagrams at $n^{\rm th}$ loop order grows like $({n\over 2})!n!$ (iii) the fact that there are
no sign cancellations between graphs. 
This led Hurst to conjecture that  the analog statement holds true for other renormalizable
quantum field theories and in particular for QED. However, QED was already well-established
experimentally in 1952. To account for this fact, Hurst postulated that the QED perturbation
series, while not convergent, is an asymptotic series and thus still makes sense numerically.

Today the fact that the perturbation series is asymptotic rather than convergent is
not only well-established for QED, but believed to be a generic property of nontrivial
quantum field theories (see \cite{LeGuillou,Stevenson,Fischer,minneapolis,Zinn-Justin,tHooft}). Convergence of the series can be expected only in
trivial cases where higher-order radiative corrections are absent altogether (usually on account
of some symmetry).

In the absence of convergence, the reconstruction of the exact physical quantities from their
perturbation series must be attempted using summation methods. For this there are various
possibilities, and we mention here only the one which has been most widely used in QFT, Borel summation \cite{Zinn-Justin,tHooft}.
For a factorially divergent series 

\bear
F(\alpha)  \sim \sum_{n=0}^{\infty} c_n \alpha^{n+1}
\label{Fidv}
\ear
one defines the Borel transform as

\bear
B(t) \equiv \sum_{n=0}^{\infty} c_n {t^n\over n!}
\label{defBoreltrafo}
\ear
Assuming that $B(t)$ has no singularities on the positive real axis and does not increase too
rapidly at infinity, one can also define the Borel integral

\bear
\tilde F (\alpha) \equiv \int_0^{\infty}dt\,  e^{-{t\over\alpha}}B(t)
\label{Borelint}
\ear
$\tilde F$ is the Borel sum of the original series $F$. $F$ is asymptotic to $\tilde F$ by construction,
although the physical quantity represented by the series $F$ might still differ from $\tilde F$ by
nonperturbative terms. Even when the Borel transform has singularities it remains a
useful concept, since these singularities contain detailed information on the divergence
structure of the theory. In many cases they can be traced either to instantons (related to tunneling
between vacua) or Euclidean
bounces (related to vacuum decay) or to renormalons. Those latter typically arise
in renormalizable theories and are related to large or small loop momentum behaviour.
Diagramatically, they can be analyzed in terms of 
``infinitely long'' chains of `bubble' diagrams (for a review, see \cite{beneke}).

Despite of the many insights which have been gained along these lines, a point which
remains poorly understood is the influence of gauge cancellations on the divergence
structure of a gauge theory. Generally, in gauge theory individual diagrams do not
give gauge invariant results; gauge invariance is recovered only after summing over
certain classes of diagrams. In QED a textbook example is provided by one-loop
photon-photon scattering, where a gauge invariant (as well as UV finite) result
is obtained only after performing a sum over the six inequivalent orderings of
the external momenta for the basic diagram in fig. 1.

\begin{picture}(20000,11000)(1000,-1000)
\put(5000,5000){\circle{5000}}
\put(13800,5000){\circle{5000}}
\drawline\photon[\NW\REG](3600,6400)[3]
\drawline\photon[\NE\REG](6400,6400)[3]
\drawline\photon[\SW\REG](3600,3600)[3]
\drawline\photon[\SE\REG](6400,3600)[3]
\put(9200,4800){\large +}
\drawline\photon[\NW\REG](12400,6400)[3]
\drawline\photon[\NE\REG](15200,6400)[3]
\drawline\photon[\SW\REG](12400,3600)[3]
\drawline\photon[\SE\REG](15200,3600)[3]
\put(17600,4800){\large +}
\put(19300,5000){\large \ldots}
\put(1000,8000){1}
\put(8300,8000){2}
\put(8300,1200){3}
\put(1000,1200){4}
\put(9800,8000){1}
\put(17100,8000){2}
\put(17100,1200){4}
\put(9800,1200){3}
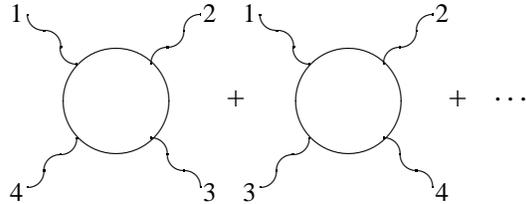
\put(1000,-1000){Fig. 1: Sum of one loop photon scattering diagrams.}
\end{picture}

\vspace{5pt}
This recovery of gauge invariance generally implies cancellations in the sum 
over gauge related diagrams, and leads
one to expect that the coefficients of the perturbation series for such amplitudes should 
come out smaller in magnitude
than predicted by a naive combinatorial analysis. However, it is presently not known
what implications this has for the large order behaviour of amplitudes in either
QED or other gauge theories.
In 1977 Cvitanovic \cite{cvitanovic} performed a detailed diagramatic analysis 
of the effect of gauge cancellations in a previous calculation of the sixth order
contribution to the electron magnetic moment \cite{cvikin}.  Based on this example,
he suggested that in QED their effect is sufficiently strong to modify
estimates based on the basic factorial growth in the number of Feynman diagrams;
more realistic estimates might be obtained by counting gauge invariant classes of
diagrams instead. The asymptotic growth of the number of these classes is, however, 
less than factorial.   
For the case at hand, the electron $g-2$, he conjectured that its perturbation series 
even {\sl converges}  in the `quenched' approximation, i.e. for the contribution
represented by diagrams not involving electron loops.       

In the present paper, we present an analogous conjecture for the $N$ - photon amplitudes,
in scalar and spinor QED. Here the `quenched' approximation corresponds to taking
only the diagrams involving just one electron loop, which is also the
O$(N_f)$ part of the amplitude in QED with $N_f$ flavors.
Our conjecture is that this
part of the amplitude converges in perturbation theory when renormalized on-shell.
We use known results on the QED effective Lagrangian to obtain information 
on the large order behaviour of this amplitude for the special case of   
`all +' polarizations, in the limit of low photon energies and
large photon number $N$.
     
\vspace{-5pt}     
\section{The Euler-Heisenberg Lagrangian and the $N$ - photon amplitudes: 
general case}
\label{eh}
\renewcommand{\theequation}{2.\arabic{equation}}
\setcounter{equation}{0}

\vspace{-2pt}

Let us start with some basic facts about the QED effective Lagrangian in a constant
external field $F_{\mu\nu}$. At one loop in spinor QED, this is the
well-known Euler-Heisenberg (`EH') Lagrangian, obtained 1936 by Heisenberg and Euler \cite{eh}
in form of the following integral:

\bear
{\cal L}^{(1)}_{\rm spin}(F)&=& - {1\over 8\pi^2}
\int_0^{\infty}{dT\over T^3}
\,\e^{-m^2T} 
\biggl[
{(eaT)(ebT)\over \tanh(eaT)\tan(ebT)} 
\nonumber\\&&\hspace{70pt}
- {1\over 3}(a^2-b^2)T^2 -1
\biggr]
\label{eh1spin}
\ear
Here $T$ is the proper-time of the loop fermion, $m$ its mass, and $a,b$ are the two
Maxwell invariants, related to $\bf E$, $\bf B$ by $a^2-b^2 = B^2-E^2,\quad ab = {\bf E}\cdot {\bf B}$.
A similar representation exists for scalar QED \cite{weisskopf,schwinger51}:

\bear
{\cal L}^{(1)}_{\rm scal}(F)&=&  {1\over 16\pi^2}
\int_0^{\infty}{dT\over T^3}
\,\e^{-m^2T} 
\biggl[
{(eaT)(ebT)\over \sinh(eaT)\sin(ebT)} 
\nonumber\\&&\hspace{60pt}
+{1\over 6}(a^2-b^2)T^2 -1
\biggr]
\label{eh1scal}
\ear

By standard QFT (see, e.g., \cite{itzzub}) these effective Lagrangians contain the information
on the one-loop photon S-matrix in the limit where all photon energies can be neglected compared
to the electron mass, $\omega_i << m$. Thus diagramatically
${\cal L}_{\rm spin}^{(1)}(F)$ is equivalent to the sum of the Feynman graphs in fig. 2 at zero momentum.

\begin{picture}(20000,13000)(1000,-3000)
\put(3000,4000){\circle{3000}}
\put(9000,4000){\circle{3000}}
\put(16000,4000){\circle{3000}}
\drawline\photon[\NW\REG](2000,5000)[2]
\drawline\photon[\NE\REG](4000,5000)[2]
\drawline\photon[\SW\REG](2000,3000)[2]
\drawline\photon[\SE\REG](4000,3000)[2]
\put(5600,3600){\large +}
\drawline\photon[\NW\REG](8000,5000)[2]
\drawline\photon[\NE\REG](10000,5000)[2]
\drawline\photon[\SW\REG](8000,3000)[2]
\drawline\photon[\SE\REG](10000,3000)[2]
\drawline\photon[\N\REG](9000,5400)[2]
\drawline\photon[\S\REG](9000,2600)[2]
\put(11100,3600){\large +}
\drawline\photon[\NW\REG](15000,5000)[2]
\drawline\photon[\NE\REG](17000,5000)[2]
\drawline\photon[\SW\REG](15000,3000)[2]
\drawline\photon[\SE\REG](17000,3000)[2]
\drawline\photon[\N\REG](16000,5400)[2]
\drawline\photon[\S\REG](16000,2600)[2]
\drawline\photon[\W\REG](14600,4000)[2]
\drawline\photon[\E\REG](17400,4000)[2]
\put(20000,3600){\large +}
\put(21000,3900){\large \ldots}
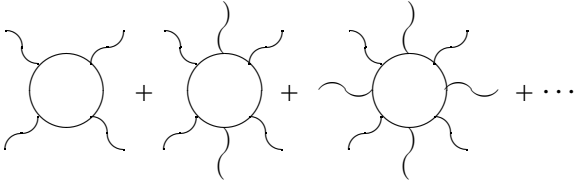
\put(1000,-2000){Fig. 2: Diagrams equivalent to the EH Lagrangian.}
\end{picture}

To obtain the one-loop $N$ - photon amplitude 
from the Lagrangian, introduce for each photon leg the field strength tensor 

\bear
F_i^{\mu\nu}= k_i^{\mu}\varepsilon_i^{\nu}-k_i^{\nu}\varepsilon_i^{\mu}
\label{defFi}
\ear
with $\varepsilon_i$, $k_i$ the photon polarization and momentum. Define

\bear
F_{\rm total} = \sum_{i=1}^N  F_i \quad
\label{defFtotal}
\ear
Then 

\bear
\Gamma^{(1)}[k_1,\varepsilon_1;\ldots;k_N,\varepsilon_N]
=
{\cal L}^{(1)}(iF_{\rm total})\mid_{F_1\cdots F_N}
\label{GammatoL}
\ear
On the right hand side it is understood that, after expanding out 
${\cal L}^{(1)}(iF_{\rm total})$ to the $N$th order in the field, only those terms
are retained which involve each $F_1,\ldots,F_N$ linearly.
Using this expansion together with a convenient choice of 
polarizations, one can obtain a closed-form expression for 
the low energy limit of the one-loop on-shell photon amplitudes, valid 
for arbitrary $N$ and all polarization components \cite{dsjhep1,nphot}.
We will be concerned here only with the `all $+$' or `maximally helicity violating' 
component of the amplitude, defined by choosing the polarization
vectors for all photons as positive helicity  eigenstates
(the `all $+$' and `all $-$' components are related by  a parity transformation). 
This component becomes particularly simple:

\bear
\Gamma_{\rm spin}^{(1)}
[k_1,\varepsilon_1^+;\ldots ;k_N,\varepsilon_N^+]
\!\!\!&=& \!\!\!
-2\frac{(2e)^{N}}{(4\pi)^2m^{2N-4}}\,c_{\rm spin}^{(1)}
({\scriptstyle\frac{N}{2}})
\chi_N \non\\
c_{\rm spin}^{(1)}(n)&=& 
- \frac{{\cal B}_{2n}}{2n(2n-2)}
\non\\
\label{Gamma1loop}
\ear
Here the ${\cal B}_k$ are the Bernoulli numbers, and $\chi_N$ is a kinematical
factor which in spinor helicity notation (see, e.g., \cite{dixon}) can be written as

\bear
\chi_N \!\!\!\!
&=& \!\!\!\!
{({\scriptstyle \frac{N}{2}})!
\over 2^{N\over 2}}
\Bigl\lbrace
[12]^2[34]^2\cdots [(N-1)N]^2 + {\rm  permutations}
\Bigr\rbrace\non\\
\label{defchiN}
\ear
($[ij]=\langle k_i^+\vert k_j^-\rangle$). Its precise form is not essential for our
purpose. Here and in the following it should be understood that we deal
with the low energy limits of the amplitudes only.

\section{Two loop: the self-dual case}
\label{twoloop}
\renewcommand{\theequation}{3.\arabic{equation}}
\setcounter{equation}{0}
The two-loop correction ${\cal L}^{(2)}(F)$ to the Euler--Heisenberg Lagrangian, involving
an additional photon exchange in the loop, has been considered by various authors starting 
with V.I. Ritus in 1975 \cite{ritusspin,lebrit,ditreuqed,rss}. However, for a general constant field
$F_{\mu\nu}$ no representation is known for ${\cal L}^{(2)}(F)$ 
which would be explicit enough to derive closed all-$N$ formulas for the two-loop
photon amplitudes. Things simplify dramatically, though, if one specializes the
field to a (Euclidean) self dual (`SD') field, obeying

\bear
F_{\mu\nu} = \tilde F_{\mu\nu} \equiv {1\over 2} \varepsilon_{\mu\nu\alpha\beta}F^{\alpha\beta}
\label{condsd}
\ear
The square of $F$ becomes then proportional to the identity matrix, $F^2 = -f^2\Eins$,
and we can use the number $f$ to parametrize $F$. 
$f$ can be either real or imaginary. The real $f$ case corresponds to a field which in 
Minkowski space has a real magnetic component and an imaginary electric component.
In Euclidean space (with $x_4=it$) the field strength matrix can be Lorentz transformed to the form

\bear
F\quad =\quad
\left(
\begin{array}{*{4}{c}}
0&f&0&0\\
-f&0&0&0\\
0&0&0&f\\
0&0&-f&0\\
\end{array}
\right)
\ear
\vspace{5pt}

\noindent
Thus it describes a ``doubling up'' of the magnetic case, and indeed 
the EH Lagrangian for such a field turns out to have 
the same qualitative properties as the EH
Lagrangian for a purely magnetic field. The self-dual case with
a real $f$ will therefore be called `magnetic' in the following.
Similarly, for a self dual field with a purely imaginary $f$ the properties
of the Lagrangian are similar to the electric EH Lagrangian and
we call this case `electric'.

In this self dual case all integrals turn out to be elementary, leading to the following simple
closed-form expression for the two-loop effective Lagrangian \cite{dsjhep1},

\bear
{\cal L}_{\rm spin}^{(2)}(\kappa)
&=&
-2\alpha \,{m^4\over (4\pi)^3}\frac{1}{\kappa^2}\left[
3\xi^2 (\kappa)
-\xi'(\kappa)\right]
\nonumber\\
\label{LspinSD2l}
\ear\no
Here $\kappa\equiv \frac{m^2}{2e\sqrt{f^2}}$ and

\bear
\xi(x)\equiv -x\Bigl(\psi(x)-\ln(x)+{1\over 2x}\Bigr)
\label{defxi}
\ear
where $\psi(x)=\Gamma'(x)/\Gamma(x)$ is the digamma function.
Since self dual fields are helicity eigenstates \cite{dufish} the effective
action for such fields carries precisely the information on the `all $+$' 
(or `all $-$') photon amplitudes. Thus at the two-loop
level we are still able to write down a closed-form all-$N$ expression for this
particular polarization choice:

\bear
\Gamma_{\rm spin}^{(2)}
[k_1,\varepsilon_1^+;\ldots ;k_N,\varepsilon_N^+]
\!\!\!\!\!
&=&\!\!\!\!
-2\alpha\pi\frac{(2e)^{N}}{(4\pi)^2m^{2N-4}}\,c_{\rm spin}^{(2)}
({\scriptstyle{\frac{N}{2}}})
\chi_N 
\non\\
c^{(2)}_{\rm spin}(n) &=&
{1\over (2\pi)^2}\biggl\lbrace
\frac{2n-3}{2n-2}\,{\cal B}_{2n-2}
\non\\&&
\hspace{10pt}
+3\sum_{k=1}^{n-1}
{{\cal B}_{2k}\over 2k}
{{\cal B}_{2n-2k}\over (2n-2k)}
\biggr\rbrace
\non\\
\label{Gamma2loop}
\ear
(the kinematic factor $\chi_N$ is the same as in  (\ref{Gamma1loop})).

\vspace{-2pt}
\section{Higher loop orders: Im$\,{\cal L}$}
\label{multiloop}
\renewcommand{\theequation}{4.\arabic{equation}}
\setcounter{equation}{0}
\vspace{-2pt}

Beyond two loops, it gets rather hard to obtain information on the
EH Lagrangian by a direct calculation. Nevertheless,
the work of Ritus and Lebedev quoted above \cite{ritusspin,lebrit} leads to
an all-order prediction for the {\sl imaginary part} of this Lagrangian.
As is well-known, the EH Lagrangian will have an imaginary part whenever
the field is not purely magnetic. Physically it represents the possibility of vacuum
pair creation by the electric field component. At one loop for the case of a purely
electric field, Schwinger \cite{schwinger51} found for it the following representation,

\bear
{\rm Im} {\cal L}_{\rm spin}^{(1)}(E) &=&  \frac{m^4}{8\pi^3}
\beta^2\, \sum_{k=1}^\infty \frac{1}{k^2}
\,\exp\left[-\frac{\pi k}{\beta}\right]
\label{ImL1}
\ear
with $\beta = {eE\over m^2}$. Here the term with index $k$ describes the coherent
production of $k$ pairs by the field. This representation is manifestly nonperturbative
in the field and coupling, making it clear that the imaginary part cannot be seen
unless one sums up the whole set of diagrams in fig. 2. 

At the two-loop level, the imaginary part of ${\cal L}(E)$ was first studied by
Lebedev and Ritus \cite{lebrit} who found the following generalization of
Schwinger's formula (\ref{ImL1}):

\bear
{\rm Im} {\cal L}_{\rm spin}^{(2)} (E) &=&  \frac{m^4}{8\pi^3}
\beta^2\,
\sum_{k=1}^\infty
\alpha\pi K_k(\beta)
\,\exp\left[-\frac{\pi k}{\beta}\right]
\non\\
\label{ImL2}
\ear
Here $\alpha=\frac{e^2}{4\pi}$ is the fine-structure constant. The exponential
factors are the same as in (\ref{ImL1}), however at two loop the $k$th exponential
comes with a prefactor $K_{k}(\beta)$ which is itself a function of the field. 
These prefactor functions are not known explicitly, but \cite{lebrit}
were able to show that they have small $\beta$ expansions of the following form:

\begin{eqnarray}
K_k(\beta) &=& -{c_k\over \sqrt{\beta}} + 1 + {\rm O}(\sqrt{\beta})
 \nonumber\\
c_1 = 0,\quad && \quad
c_k = {1\over 2\sqrt{k}}
\sum_{l=1}^{k-1} {1\over \sqrt{l(k-l)}},
\quad k \geq 2\non\\
\label{expK}
\end{eqnarray}
In the following we will concentrate on the weak field limit $\beta <<1$ of ${\rm Im}{\cal L}$. 
In this limit the $k$ - series in (\ref{ImL1}), (\ref{ImL2}) are dominated by the $k=1$ term, and
from (\ref{expK}) ${\rm lim}_{\beta\to 0}K_1(\beta) = 1$. Thus one has

\bear
{\rm Im} {\cal L}_{\rm spin}^{(1)} (E) +
{\rm Im}{\cal L}_{\rm spin}^{(2)} (E) 
\,\,\,\, {\stackrel{\beta\to 0}{\sim}} \,\,\,\,
 \frac{m^4\beta^2}{8\pi^3}
\bigl(1+\alpha\pi\bigr)
\,{\rm e}^{-{\pi\over\beta}}
\non\\
\label{Im1plus2}
\ear
Now \cite{lebrit} find that the $\alpha\pi$ term in (\ref{Im1plus2}) has a natural
interpetation in the pair creation picture. Naively, to turn real a virtual electron-positron
must separate out along the field direction a distance $r_{\parallel}$ such that the energy
gained from the field makes up for the rest masses. This is

\bear
r_{\parallel} = {2m\over eE}
\label{rparnaive}
\ear
However, this does not take into account the Coulomb attraction; a pair getting 
``born'' at a separation of $r_{\parallel}$ comes with a negative binding
energy of $-\alpha/ r_{\parallel}$.
Lebedev and Ritus argue that, since this energy reduces the amount of energy which
has to be taken out of the field, it is equivalent to a lowering of the rest mass. Thus
$m$, the physical (on-shell renormalized) vacuum mass, should be replaced by an
effective field-dependent mass $m_*(E)$,

\bear
m_*(E) &=& m - {\alpha\over 2}{eE\over m}
\label{defmstar}
\ear
with corrections expected to be of higher order in $\beta$ (see \cite{ritusmass,ginzburg} for
a detailed discussion). Then replacing $m$ by $m_*(E)$ in the one-loop $k=1$ exponential
and expanding in $\alpha$ one finds that

\bear
{\rm exp}\biggl[-\pi {m_*^2(E)\over eE}
\biggr]
= 
\Bigl[1 + \alpha\pi + {\rm O}(\alpha^2)\Bigr]
{\rm exp}
\biggl[
-\pi {m^2\over eE}
\biggr]\non\\
\label{expexp}
\ear
Thus if this interpretation is correct then the $1+\alpha\pi$ in the two-loop formula
(\ref{Im1plus2}) is the truncation of an exponential series, and the missing higher order
terms must show up in the higher loop corrections to ${\rm Im}{\cal L}_{\rm spin}(E)$.
Moreover, in the weak field limit these should be the {\sl only} missing terms, leading to
a remarkably simple all-order prediction:

\bear
\sum_{l=1}^{\infty}{\rm Im}{\cal L}^{(l)}_{\rm spin}(E)
&{\stackrel{\beta\to 0}{\sim}}&
 \frac{m^4\beta^2}{8\pi^3}
\,{\rm exp}\Bigl[ -{\pi\over\beta}+\alpha\pi \Bigr]
\label{ImLallloop}
\ear
In the weak field limit the above considerations are 
spin-independent \cite{ginzburg} so that the same 
formula (\ref{ImLallloop}) with just an additional
factor of $1/2$ on the right hand side applies to scalar QED. For this
case, moreover, there exists a completely different and more direct
derivation of the same formula due to Affleck et al. \cite{afalma}.
(These derivations were independent of one another: apparently Affleck et al. were 
not aware of Ritus's work, and Ritus and Lebedev were not aware of Affleck et al.'s work.) 
It uses the following `worldline path integral' representation of the quenched effective action
in scalar QED \cite{feynman50},

\bear
\Gamma_{\rm scal}^{({\rm quenched})}(A)
&=&
\int_0^{\infty}{dT\over T}\,{\rm e}^{-m^2T}
\int{\cal D}x(\tau)
\, e^{-S[x(\tau)]}
\nonumber\\
\label{wlpi}
\ear
Here the 
path integral ${\cal D}x(\tau)$ runs over the space of all
embeddings $x(\tau)$ of the particle trajectory into
(Euclidean) spacetime, with periodic boundary conditions
in proper-time $x(T) = x(0)$. The path integral
action $S[x(\tau)]$ has three parts,

\bear
S=S_0+S_e+S_i
\label{Ssum}
\ear

\bear
S_0 &=& \int_0^T d\tau {\dot x^2\over 4}  \nonumber\\
S_e &=& ie\int_0^T \dot x^{\mu}A_{\mu}(x(\tau))
\nonumber\\
S_i &=&
-{e^2\over 8\pi^2}\int_0^Td\tau_1\int_0^Td\tau_2 {\dot x(\tau_1)\cdot\dot x(\tau_2)\over
(x(\tau_1)-x(\tau_2))^2}\nonumber
\non\\
\label{Sparts}
\ear
$S_e$ describes the interaction of the scalar with the external field, $S_i$ 
the exchange of internal photons in the scalar loop. In \cite{afalma} this
representation was applied to the constant electric field case, and
Im${\cal L}_{\rm scal}(E)$ was calculated using a stationary path approximation
both for the $T$ integral and the path integral $\int {\cal D}x(\tau)$. The stationary
trajectory turns out to be a circle with a field dependent radius. Evaluation of the
worldline action $S[x(\tau)]$ on this trajectory yielded exactly the exponent in
(\ref{ImLallloop}), and the second variation determinant gave precisely the same prefactor.
Affleck et al. then argued that the stationary point approximation becomes exact
in the weak field limit up to renormalization effects.

Formulas analogous to (\ref{ImL1}), (\ref{ImL2}) exist 
also for the self dual case \cite{dsjhep2}. In the `magnetic' case
(real $f$) the EH Lagrangian is real, while in the `electric' case (imaginary $f$) 
it has an imaginary part. At one and two loops, this imaginary part has the
following Schwinger type expansion,

\bear
{\rm Im} {\cal
L}^{(1)}_{\rm spin}(i\kappa)
&=&-\frac{m^4}{(4\pi)^2}\frac{1}{\kappa}\sum_{k=1}^\infty
\left(\frac{1}{k}+\frac{1}{2\pi \kappa k^2}\right)\, e^{-2\pi k \kappa}
\non\\
{\rm Im}{\cal L}^{(2)}_{\rm spin}(i\kappa) &=&
-\alpha \pi
\frac{m^4}{(4\pi)^2}
\frac{1}{\kappa}\sum_{k=1}^\infty K_k(\kappa)
\, e^{-2\pi\kappa k}
\non\\
\label{ImLdsjhep12}
\ear
Here the functions $K_k(\kappa)$ are the SD analogues of the 
Lebedev--Ritus functions $K_k(\beta)$. In contrast to the
$K_k(\beta)$, the $K_k(\kappa)$ are known explicitly as 
power series in $1/{\kappa}$ \cite{moscow}: 

\bear
K_k(\kappa) &=&
k-\frac{1}{2\pi\kappa}-
\frac{3\kappa}{\pi} \sum_{m=1}^\infty
\frac{(-1)^m{\cal B}_{2m}}{2m{\kappa}^{2m}}
\nonumber\\
\label{KkSD}
\ear
In the weak field limit $f\to 0$ or $\kappa\to\infty$ one finds the same relation between
the one and two loop contributions to Im${\cal L}$ as in (\ref{Im1plus2}),

\bear
{\rm Im}{\cal L}^{(2)}_{\rm spin}(i\kappa)
\quad {\stackrel{f\to 0}{\sim}}\quad
\alpha\pi\, {\rm Im}{\cal L}^{(1)}_{\rm spin}(i\kappa)
\label{L2toL1SD}
\ear
Although neither the Lebedev-Ritus arguments nor the approach of
Affleck et al. carry over to the self dual case in an obvious way, we
take this as evidence that the simple exponentiation 
(\ref{ImLallloop}) applies also to the self dual case. Thus we conjecture
that

\bear
\sum_{l=1}^{\infty}{\rm Im}{\cal L}^{(l)}_{\rm spin}(i\kappa)
&{\stackrel{f\to 0}{\sim}}&
-\frac{m^4}{(4\pi)^2}\frac{1}{\kappa}
\, {\rm exp}\Bigl[-2\pi \kappa +\alpha\pi \Bigr]\non\\
\label{ImLallloopSD}
\ear

\vspace{-20pt}
\section{Re${\cal L}\leftrightarrow\,$Im${\cal L}$ via Borel dispersion relations}
\label{borel}
\renewcommand{\theequation}{5.\arabic{equation}}
\setcounter{equation}{0}
\vspace{-2pt}

We would now like to convert the result (\ref{ImLallloop}) for Im$\cal L$ into 
information about Re$\cal L$ and the photon amplitudes. The appropriate
dispersion relation involves the same Borel technique which we described
in the introduction for the loop expansion, now 
applied to the weak field expansion.
Let us first demonstrate it at the one loop level \cite{Chadha}. In the purely magnetic
case, it is easy to obtain from (\ref{eh1spin}) the following closed formula for the
coefficients of the weak field expansion of ${\cal L}_{\rm spin}^{(1)}$,

\bear
{\cal L}_{\rm spin}^{(1)}(B)
&=&
{2m^4\over\pi^2}
\sum_{n=0}^{\infty}
a_n^{(1)}g^{n+2}
\label{Lmagexp}
\ear
where $g=\Bigl({eB\over m^2}\Bigr)^2$ and 

\bear
a_n^{(1)} = -{2^{2n}{\cal B}_{2n+4}\over (2n+4)(2n+3)(2n+2)}
\label{coeffLmag}
\ear
Using properties of the Bernoulli numbers one can then show
that asymptotically the $a_n^{(1)}$ behave as

\bear
a_n^{(1)}\,\,
{\stackrel {n\to\infty} \sim} \quad \!\!
{(-1)^n \over 8\pi^4}{\Gamma(2n+2)\over\pi^{2n}}
\Bigl(1\!+\!{1\over 2^{2n+4}}\!+\!{1\over 3^{2n+4}}\!+ \ldots \Bigr)
\non\\
\label{coeffLmagasymp}
\ear
If one replaces $a_n^{(1)}$ with just the leading term in this expansion, the
series (\ref{Lmagexp}) turns into an alternating series which is divergent but Borel
summable, i.e. the Borel integral (\ref{Borelint}) is well-defined. The weak field expansion
of ${\cal L}_{\rm spin}^{(1)}$ for the purely electric case differs from (\ref{Lmagexp}) 
only by an additional factor $(-1)^n$:

\vspace{-4pt}
\bear
{\cal L}_{\rm spin}^{(1)}(E)
&=&
{2m^4\over\pi^2}
\sum_{n=0}^{\infty}
(-1)^n a_n^{(1)}g^{n+2}
\label{Lelecexp}
\ear
where now $g=\Bigl({eE\over m^2}\Bigr)^2$. In this case the same leading approximation
produces a non-alternating series for which the Borel integral diverges. Nevertheless,
the imaginary part of this integral is well-defined through a dispersion relation (using
the discontinuity across the cut along the negative g axis) and yields just the 
$k=1$ term in the Schwinger expansion of Im${\cal L}_{\rm spin}^{(1)}$, eq. (\ref{ImL1}).
Applying the same procedure to the whole asymptotic expansion (\ref{coeffLmagasymp})
term by term one reproduces the complete Schwinger expansion. 

At the two loop level, no closed formula is known for the corresponding coefficients
$a_n^{(2)}$. Still, in \cite{ds1} their leading and subleading asymptotic
growth could be determined by calculating the first fourteen coefficients and using them
for a numerical fit. This yielded 

\bear
a_n^{(2)} \quad
{\stackrel {n\to\infty} \sim }
\quad \alpha{(-1)^n \over 8\pi^3}{\Gamma(2n+2)\over\pi^{2n}}
\Bigl(1 - {0.44\over \sqrt{n}}  + \ldots \Bigr)
\label{coeffLmagasymp2}
\ear
The leading term differs only by a factor $\alpha\pi$ from the corresponding one loop
term, and thus when used in the Borel dispersion relation gives the $\alpha\pi$ term
in (\ref{Im1plus2}). From the subleading term one obtains the second term in the
small $\beta$ expansion (\ref{expK}) for $K_1$,

\vspace{-9pt}
\bear
K_1(\beta) = 1 - ( 0.44)\sqrt{{2\over\pi}\beta} + {\rm O}(\beta) 
\label{K1exp}
\ear 
Thus the Borel technique allows one to map the asymptotic large $n$ expansion of the 
weak field expansion coefficients $a_n$ of the magnetic Lagrangian to the weak field 
expansion of the imaginary part of the electric Lagrangian.

In the self dual case, where closed formulas are available for the real and imaginary parts
of the EH Lagrangian also at two loops, this correspondence can be checked in even
more detail. In \cite{dsjhep2} we calculated, up to exponentially suppressed terms, the large $n$ expansion
of the two loop weak field expansion coefficients 
$c_{\rm scal}^{(2)}(n)$ for scalar QED (the $c_{\rm scal}^{(2)}(n)$ differ from the 
$c_{\rm spin}^{(2)}(n)$ given in (\ref{Gamma2loop}) only in the coefficient of the second term in braces,
which in the scalar QED case is $3/2$ instead of $3$). Using a Borel dispersion relation as above 
then yielded precisely the small $f$ expansion of (the scalar QED equivalent of) $K_1(\kappa)$.

\vspace{-3pt}
\section{Large $N$ behaviour of the $l$-loop $N$- photon amplitudes}
\label{largeN}
\renewcommand{\theequation}{6.\arabic{equation}}
\setcounter{equation}{0}
\vspace{-5pt}

Assuming that this Borel dispersion relation remains valid at higher loop orders,
we can apply it in reverse to the exponentiation formula
(\ref{ImLallloopSD}) and obtain all-loop information on the coefficients of the weak field
expansion:

\vspace{-8pt}
\bear
{{\rm lim}_{n\to\infty}} {c^{(l)}_{\rm spin}(n)\over c^{(1)}_{\rm spin}(n)} = {(\alpha\pi)^{l-1}\over (l-1)!}
\label{limratiocoeff}
\ear
\vspace{-3pt}

\noindent
By (\ref{GammatoL}) this translates into a statement on the `all +' amplitudes
in the limit of large photon number $N$,

\bear
{\rm lim}_{N\to\infty}{\Gamma_{\rm spin}^{(l)}
[k_1,\varepsilon_1^+;\ldots ;k_N,\varepsilon_N^+]
\over
\Gamma_{\rm spin}^{(1)}
[k_1,\varepsilon_1^+;\ldots ;k_N,\varepsilon_N^+]
} = {(\alpha\pi)^{l-1}\over (l-1)!}
\label{limratioamp}
\ear
Summing this relation over $l$ we formally get an exponentiation
formula for the complete amplitude in the large $N$ limit:

\bear
{\rm lim}_{N\to\infty}{\Gamma_{\rm spin}^{({\rm total})}
[k_1,\varepsilon_1^+;\ldots ;k_N,\varepsilon_N^+]
\over 
\Gamma_{\rm spin}^{(1)}
[k_1,\varepsilon_1^+;\ldots ;k_N,\varepsilon_N^+]}
=
\e^{\alpha\pi}
\non\\
\label{ampexp}
\ear
Assuming sufficient uniformity in $l$ of the convergence of the ratio (\ref{limratioamp}) for $N\to\infty$, one could now conclude that the amplitude must be analytic in $\alpha$ for some
sufficiently large $N$. But analyticity of the complete amplitude is excluded by renormalons and 
other arguments. Therefore uniformity must fail, and it is easy to see how this comes about
diagramatically. Fig.  3 shows the  Feynman diagrams for the $N$ photon amplitude for the first four
loop orders (the external photons are not displayed, as well as diagrams differing from those
depicted only by moving the end points of photon lines along the fermion loops). 
At $l$ loops, it involves diagrams with up to $l-1$ fermion loops. 

\begin{picture}(20000,40000)(1000,-3000)
\put(0,32700){l=1}
\put(5000,33000){\circle{3050}}
\put(0,27700){l=2}
\put(5000,28000){\circle{3050}}
\drawline\photon[\N\REG](5000,26420)[3]
\put(0,22700){l=3}
\put(5000,23000){\circle{3050}}
\drawline\photon[\N\REG](3850,22020)[2]
\drawline\photon[\N\REG](6200,22020)[2]
\put(12000,23000){\circle{3050}}
\put(16600,23000){\circle{3050}}
\drawline\photon[\E\REG](13400,22000)[2]
\drawline\photon[\E\REG](13400,24000)[2]
\put(0,17700){l=4}
\put(5000,18000){\circle{3050}}
\drawline\photon[\N\REG](5000,16420)[3]
\drawline\photon[\N\REG](3850,17020)[2]
\drawline\photon[\N\REG](6200,17020)[2]

\put(5000,13000){\circle{3050}}
\put(10230,13000){\circle{3050}}
\drawline\photon[\E\REG](6250,14200)[3]
\drawline\photon[\E\REG](6250,11800)[3]
\drawline\photon[\E\REG](6660,13000)[2]

\put(15000,13000){\circle{3050}}
\put(19600,13000){\circle{3050}}
\drawline\photon[\E\REG](16400,12000)[2]
\drawline\photon[\E\REG](16400,14000)[2]
\drawline\photon[\N\REG](15000,11420)[3]

\put(10000,4000){\circle{3050}}
\put(15230,4000){\circle{3050}}
\put(12615,8040){\circle{3050}}
\drawline\photon[\NE\REG](11000,5240)[2]
\drawline\photon[\NW\REG](14230,5240)[2]
\drawline\photon[\E\REG](11660,4000)[2]

\put(0,1000){\vdots}

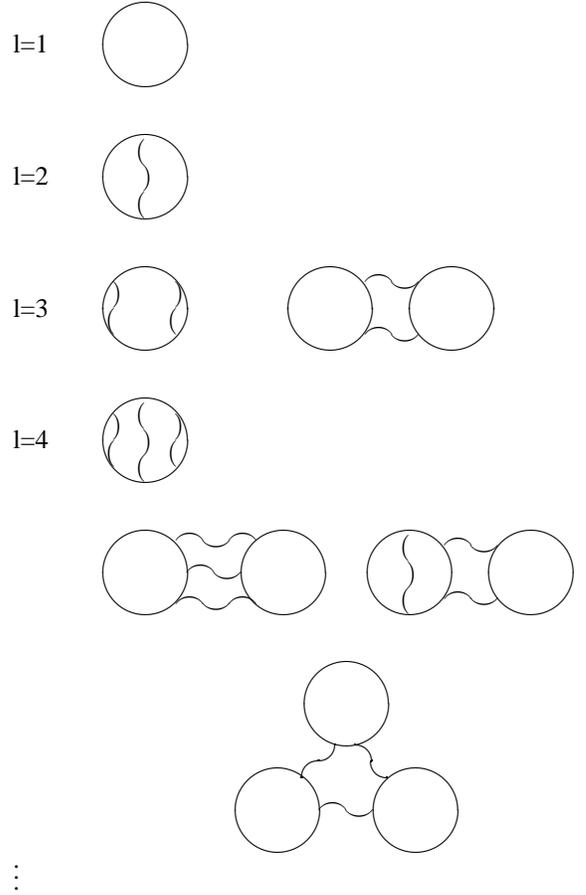
\put(0,-2000){Fig. 3: Feynman diagrams up to four loop order.}

\end{picture}

From the Affleck et al. approach to the derivation of the exponentiation formula (\ref{ImLallloop}) 
we know that the weak field limit of Im${\cal L}(E)$ comes entirely from the `quenched' part of
the amplitude. Thus at fixed loop order $l$ the total contribution of all non-quenched diagrams
must be subleading in $N$ compared to the quenched ones. Since the number of non-quenched
classes of diagrams grows with increasing loop number, clearly one would expect the convergence
in $N$ to slow down with increasing $l$. 

If, on the other hand, one stays inside the quenched class of diagrams from the beginning, there
is no obvious mathematical reason to expect such a slowing down. Thus we believe that, for
the quenched amplitude, the perturbation series indeed converges for sufficiently large $N$,
and in fact for all $N\geq 4$;       
since it would be very surprising to find a qualitative change in the
convergence properties of these amplitudes at some definite value
of $N$ (although such a phenomenon cannot be excluded altogether,
of course).  Similarly, considering the essentially topological character of
gauge cancellations one would not assume the restriction 
to zero momenta or the choice of polarizations to be essential.
Thus we are led to the following generalization of Cvitanovic's conjecture: 
{\sl the perturbation series converges for all
on-shell renormalized QED amplitudes at leading order in $N_f$}.

It must be emphasized that on-shell renormalization is essential in all
of the above. The finiteness of the limits in (\ref{limratiocoeff}) and (\ref{limratioamp})
requires the weak field expansion coefficients $c^{(l)}_{\rm spin}(n)$ to have
the same leading asymptotic growth for all $l$, namely $\Gamma(2n-2)$ in the magnetic case
and $\Gamma(2n-1)$ in the `magnetic' self dual case. At the two loop level, one can show by
explicit calculation (numerically for the magnetic case \cite{ds1}, analytically for the `magnetic'
self dual
one \cite{dsjhep2}) that the use of any renormalized electron mass other than the on-shell
mass leads to an asymptotic growth faster than at one loop. In fact, a simple recursion
argument shows that, if mass renormalization is done generically, then the $l$ loop coefficients
will grow asymptotically like $\Gamma(2n+l-3)$ in the magnetic and like $\Gamma(2n+l-2)$ 
in the `magnetic' self dual case. 

\section{Conclusions}
\label{conclusions}
\renewcommand{\theequation}{7.\arabic{equation}}
\setcounter{equation}{0}
To summarize, we have used existing results on the imaginary part of the QED
Euler-Heisenberg Lagrangian together with Borel analysis techniques and a
number of explicit two loop checks to argue for convergence of the perturbation series
for the physically renormalized QED photon amplitudes, to leading order in $N_f$.
To the best of our knowledge, this statement is not at variance with known 
arguments against convergence for the full amplitudes, since those generally
rely on the presence of an infinite number of virtual particles. In fact, 
it presently remains an open possibility that convergence might hold 
to any finite order in $N_f$.

It would be clearly desirable to further corroborate our conjecture by higher loop
calculations. While an explicit three-loop calculation for the magnetic or electric
EH Lagrangian seems presently out of the question, for the self dual case this calculation
might be technically feasible \cite{wip}. Since the three loop amplitude has already a
non-quenched part, we can make for it the double prediction that (i) the weak field
limit of the quenched contribution to 
Im${\cal L}_{\rm spin}^{(3)}(i\kappa)$ should display the $(\alpha\pi)^2/2$ correction implicit
in (\ref{ImLallloopSD}) and (ii) the non-quenched contribution should be suppressed in
this limit. Moreover, one would like to see the ratio $c^{(3)}(n)/c^{(1)}(n)$ to converge not slower
than $c^{(2)}(n)/c^{(1)}(n)$ if only the quenched contribution is used in calculating $c^{(3)}(n)$.

\vspace{12pt}

\noindent
{\bf Acknowledgements:} We thank David Broadhurst and Dirk Kreimer for helpful
correspondence.


\end{multicols}
\medline
\begin{multicols}{2}

\end{multicols}
\end{document}